\documentclass[sigconf]{acmart}
\AtBeginDocument{%
  }

\setcopyright{none}
\acmDOI{}
\acmConference[]{}{}
\acmBooktitle{}
\acmISBN{}

\usepackage{multirow}

\newcommand{\myred}[1]{\textcolor{red}{#1}}
\newcommand{\myorange}[1]{\textcolor{orange}{#1}}  

\author{Yu Liu}
\affiliation{%
  \institution{NJUST}
  \city{Nanjing}
  \country{China}}
\email{	1iu04yui@gmail.com}

\author{Jiangxia Cao*}
\thanks{*Jiangxia Cao is the Corresponding author}
\affiliation{%
  \institution{Kuaishou Technology}
  \city{Beijing}
  \country{China}}
\email{caojiangxia@kuaishou.com}

\begin{document}

\title{Harmonizing Generative Retrieval and Ranking in Chain-of-Recommendation}


\renewcommand\footnotetextcopyrightpermission[1]{}
\settopmatter{printacmref=false}
\fancyhead{}

\begin{abstract}
Generative recommender systems have recently emerged as a promising paradigm by formulating next-item prediction as an auto-regressive semantic IDs generation, such as OneRec series works. 
However, with the next-item-agnostic prediction paradigm, its could beam out some next potential items via Semantic IDs but hard to estimate which items are better from them, e.g., select the top-10 from beam-256 items, leading to a gap between generation and ranking performance.
%
To fulfill this gap, we propose RecoChain, a unified generative retrieval and ranking framework that integrates candidate generation and ranking within a single Transformer backbone. 
Specifically, in inference, the model first generates candidate items via hierarchical semantic ID prediction, then performs the SIM-based ranking process to estimate the click possibility of corresponding item candidate continuously.
%
Extensive experiments on large-scale real-world datasets demonstrate that our approach effectively bridges the gap between generative retrieval and ranking, achieving improved Top-K recommendation performance while maintaining strong generative capability.
\end{abstract}




\maketitle

\section{Introduction}
\myorange{\textbf{Background.}}
Following the GPU-era Transformer architecture revolution in NLP/CV, the RecSys community also cast the next-item prediction task as a next-item's Semantic ID generation problem in an autoregressive manner~\cite{rajput2023recommender,zheng2024adapting}. 
This problem formulation shows a bright future in data/parameter/computation/sequencen-scaling in user behavior sequences modeling, to make the RecSys more smart~\cite{geng2022recommendation}. 
With the development of large-scale sequential data, the \myred{Semantic ID based Generative Recommenders (GR)} achieve competitive performance on next-item prediction tasks.

\myorange{\textbf{Related Work.}}
To build a powerful generative model, the item Tokenizer is a cornerstone, which aims to quantify item embeddings into several integers (e.g., 8192 * 3).
In simplify, there are several different Tokenizer variants: (1) TIGER's RQVAE~\cite{rajput2023recommender}, (2) QARM's RQ-Kmeans~\cite{luo2025qarm}, (3) QARM V2's RQ-KmeansFSQ~\cite{xia2026qarm}, (4) LLaDA-Rec's Product Parallel~\cite{shi2025llada} and so on, providing promising item quantifying mechanism.
According to the Tokenized item's Semantic ID, there are many different methods adopted in real-world applications, such as the OneRec~\cite{deng2025onerec} for short-video recommendation, OneMall~\cite{zhang2026onemall} for e-commerce recommendation and OneLive~\cite{wang2026onelive} for live-streaming recommendation.
With those elaborate works, the GRs became one of most important iteration direction in industry RecSys.

\begin{figure}[t!]
  \centering
  \includegraphics[width=9cm,height=2cm]{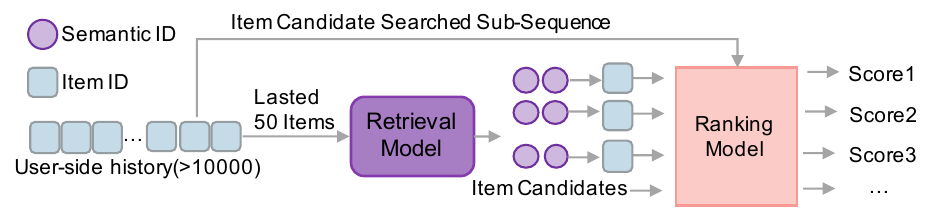}
  \caption{Two-stage Retrieval then Ranking pipeline.}
  \label{background}
\end{figure}

\myorange{\textbf{Motivation.}}
Actually, the real-world RecSys usually adopt a two-stage pipeline to give user a precise feedback~\cite{covington2016deep,liu2022neural} (as shown in Figure~\ref{background}):
\begin{itemize}
    \item Retrieval: given user-side past interested item sequence, predict a set of next item candidates, e.g., 256 different items. The problem is defined as $P(\texttt{next item}\mid\texttt{user feature})$.
    \item Ranking: based on the 256 different items, the ranking model could utilize more feature to score them one by one, and finally select the top-10 items send to users. Its problem is defined as: $P(\texttt{probability}\mid\texttt{user feature, item feature})$.
\end{itemize}
However, existing generative RecSys studies predominantly focus on the retrieval model design~\cite{rajput2023recommender,li2024large}, yet neglect the ranking process.
To alleviate the problem, those works utilize a trade-off way to distill the ranking knowledge to GR models.
The core idea is similar with the reinforcement learning, utilizing the ranking model as a reward model to teach the GR backbone which item semantic ID will be better to reach a higher positive behaviour score.
Such phenomenon showing a clear ability gap between GR and ranking, and leaves a key under-explored problem: \myred{\textit{can we unify the retrieval and ranking in one Transformer backbone}}?

\begin{figure*}[t!]
  \centering
  \includegraphics[width=17cm,height=5.3cm]{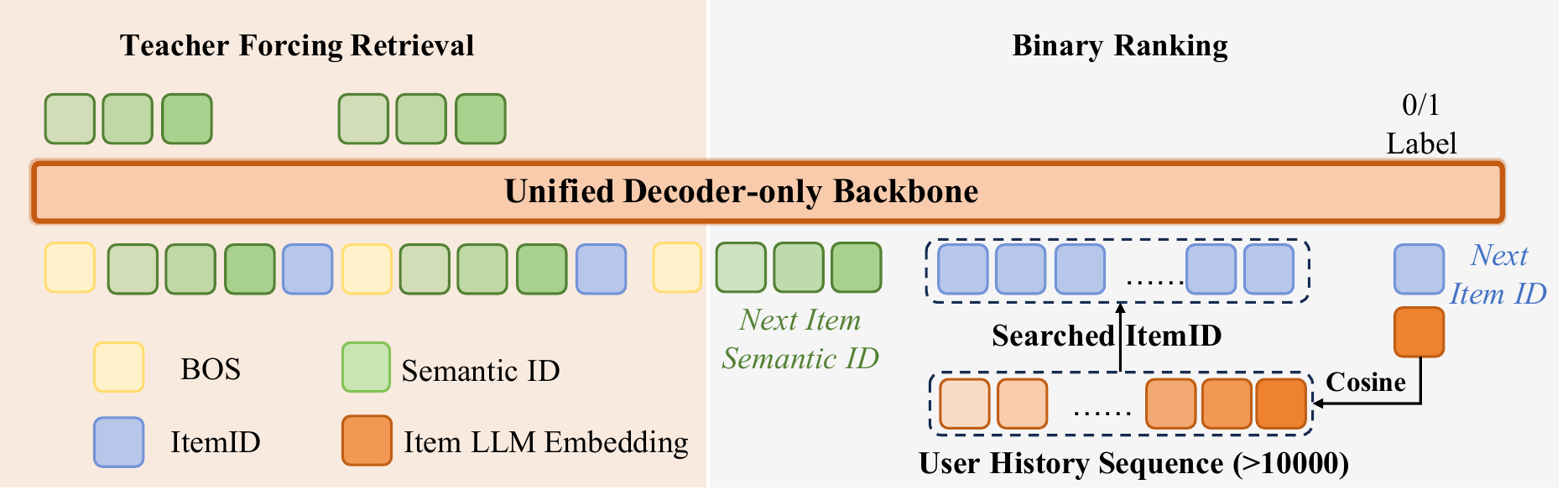}
  \caption{The training token organization of RecoChain: (1) first predict the next item Semantic ID generation, (2) then conduct the item candidate aware sub-sequence search as following input tokens to measure ranking score.}
  \label{mainmodel}
\end{figure*}

To merge the two stage in one GR model, we first revisit designing and technique routing difference among industrial retrieval/ranking model.
For the retrieval model, the research direction is focus on user latest sequence modeling and computation scaling, from the early two-tower ANN-based methods (e.g., MLP of DSSM~\cite{huang2013learning}, user-side Transformer of SASRec~\cite{kang2018self}/Kuaiformer~\cite{liu2024kuaiformer}) to the latest Beam search-based methods (e.g., TIGER~\cite{rajput2023recommender}/OneRec~\cite{deng2025onerec}).
For the ranking model, the research direction is focus on the user latest sequence modeling, target-item aware searched sequence modeling and computation scaling, from the early DIN~\cite{zhou2018deep} (user latest sequence modeling), SIM~\cite{pi2020search}/TWIN~\cite{si2024twin} (target-item aware searched sequence modeling) to the RankMixer~\cite{zhu2025rankmixer}/Wukong~\cite{zhang2024wukong}/HSTU~\cite{zhai2024actions} (computation scaling).
From those technique routing, we could find that the user latest sequence modeling and computation scaling are already included in GR models~\cite{ju2025generative}.
The major difference is that the target-item aware searched sequence modeling component is still absent in existing GR models.
As a result, GR models lack the fine-grained target-aware interaction that has proven critical for ranking accuracy, which we argue is the root cause of the ability gap between GR and ranking observed in recent distillation-based works.

\myorange{\textbf{Contribution.}}
Motivated by these observations, we propose a unified RecoChain workflow, in which a single generative backbone that conducts the retrieval and ranking in an auto-regressive manner.
Instead of treating generation and ranking as separate stages with independent models, our framework jointly performs both tasks within a unified architecture: \myred{once a candidate item (or a set of candidates) has been generated by the GR backbone, the same Transformer can be re-invoked in a target-aware mode to perform searched sequence modeling against the generated target, thereby unifying retrieval and ranking within one backbone without introducing an additional ranking component}.
In this way, the retrieval/ranking procedures could share the same KV cache and complex computation scaling unit, maximizing the training and inference efficiency.

Overall, our contributions are as follows:
\begin{itemize}
    \item We present RecoChain, a unified auto-regressive workflow, which allows a single generative model to serve both candidate generation and ranking score estimation.
    \item We conduct large-scale study on real-world TAOBAO-MM data that significantly improve the prediction accuracy.
\end{itemize}

\section{Methodology}

\subsection{Overview}
In this paper, we consider a common recommendation setting in industry: given a user history interaction sequence $$\texttt{User Log:}\ \ \{(x_n, \mathbf{m}_n), \dots, (x_1, \mathbf{m}_1), (x_0, \mathbf{m}_0)\}$$ where $x_i$ denotes item ID, $\mathbf{m}_i$ denotes item LLM multi-modal embedding.
We first conduct the Semantic ID based retrieval as follows:
\begin{equation}
\begin{split}
\texttt{Next}\ (x, \textbf{m}) \leftarrow{} \texttt{Backbone}\big((x_L, \mathbf{m}_L), \dots, (x_1, \mathbf{m}_1), (x_0, \mathbf{m}_0)\big)
\end{split}
\label{grgr}
\end{equation}
where $L$ is a small value (e.g., 50) to represent the latest item sequence.
Based on the generated next item candidate $x$, we first conduct the GSU search to retrieve the related sub sequence:
\begin{equation}
\small
\{(x^s_g, \mathbf{m}^s_g), \dots, (x^s_0, \mathbf{m}^s_0)\}
= 
\texttt{Cos-Sim}\big(\mathbf{m}, \{(x_n, \mathbf{m}_n), \dots, (x_0, \mathbf{m}_0)\}\big)
\label{grrank}
\end{equation}
where the \texttt{Cos-Sim} denote the cosine similarity function to retrieve other similar items from multi-modal embedding $\mathbf{m}$ space, $g$ is a hyper-parameter to control the searched sub-sequence length.
Finally, we feed the searched sequence to the same backbone to estimate binary 0/1 ranking score.

\begin{equation}
\begin{split}
\text{RankScore} \leftarrow \texttt{Backbone}\Big(
& \{(x_L,\mathbf{m}_L),\dots,(x_0,\mathbf{m}_0)\}, \\
& \{(x_g^s,\mathbf{m}_g^s),\dots,(x_0^s,\mathbf{m}_0^s)\},\ x
\Big).
\end{split}
\label{eq:grrank}
\end{equation}

In this way, we could utilize the \texttt{Rank\_Score} to sort different item candidates, as shown in Figure~\ref{mainmodel}.




\subsection{Item Tokenizer}


To generate the item-side Semantic ID, enable generative modeling over a large item space, we represent each item using a hierarchical semantic ID:
\begin{equation}
\begin{split}
(s^1, s^2, s^3) \leftarrow{} \texttt{RQ-Kmeans}(\mathbf{m})
\end{split}
\label{tokenizer}
\end{equation}
where the $(s^1, s^2, s^3)$ denotes the three-level Semantic ID, and we construct Semantic codebook by applying residual $k$-means quantization to the pretrained multimodal embeddings $\mathbf{m}$.
\myred{Moreover, to avoid the code conflict problem, we further append a random ID $s^4$ to ensure the one-to-one mapping between Semantic ID pattern and item ID.
Therefore, each item's Semantic ID is formed as $(s^1, s^2, s^3, s^4)$, while $s^4$ is a random integer.
}

\subsection{Sequence Modeling and Architecture}

The user history is serialized by concatenating these SID groups in chronological order.
In implementation, BOS tokens are used to mark item boundaries, while item ID tokens are additionally introduced when needed for candidate-specific matching; these serialization details are illustrated in Figure~\ref{mainmodel} and omitted in the formal description for clarity.

We adopt a decoder-only Transformer as the shared backbone.
Two task-specific heads are built on top of the decoder:
(1) a SID head for next-item SID generation, and
(2) a reranking head for candidate relevance estimation.

\subsubsection{Stage I: SID Generation Pretraining}

In the first stage, the decoder is pretrained with teacher forcing to autoregressively predict the SID sequence of each item.
The supervision is defined only on SID targets, while item ID positions are excluded from loss computation.
Concretely, for each item $t$, the prediction follows the auto-regressive order
\[
\text{BOS} \rightarrow s_t^1 \rightarrow s_t^2 \rightarrow \cdots \rightarrow s_t^4 \rightarrow \text{ItemID}.
\]
After the full SID sequence is produced, an item ID token is appended in the serialized sequence as an auxiliary item-level token to enrich input information.

Let $\hat{P}(x_n \mid x_{<n})$ denote the decoder's next-token distribution at position $n$.
The Stage-I objective is defined as
\[
\mathcal{L}_{\mathrm{SID}}
=
-\sum_{t=1}^{T}\sum_{h=1}^{4}
\log \hat{P}\!\left(s_t^h \mid x_{<\,\mathrm{pos}(t,h)}\right),
\]
where $\mathrm{pos}(t,h)$ denotes the position corresponding to the prediction target $s_t^h$ in the flattened serialized sequence.

\subsubsection{Stage II: Joint Generate--Retrieve--Rerank Training}

In contrast to Stage I, which only optimizes SID generation, Stage II jointly performs candidate generation and candidate-aware reranking within a unified decoder pipeline with KV-cache reuse.

Given the serialized user history prefix $\mathcal{X}_{<T}$, the decoder first performs a forward pass to initialize the hidden states and KV cache.
Starting from this state, hierarchical beam search is applied to autoregressively generate the SID sequence of the next item:
\[
\{\tilde{\mathbf{s}}^{(c)}\}_{c=1}^{C},
\qquad
\tilde{\mathbf{s}}^{(c)} = [\tilde{s}^{(c),1}, \dots, \tilde{s}^{(c),4}].
\]
At decoding step $\ell$, the model predicts the $\ell$-th SID token for each beam while reusing the KV cache from previous steps.
For each beam candidate $c$, the generation score is
\[
\log P(\tilde{\mathbf{s}}^{(c)} \mid \mathcal{X}_{<T})
=
\sum_{\ell=1}^{4}
\log \hat{P}(\tilde{s}^{(c),\ell}\mid \tilde{\mathbf{s}}^{(c),<\ell}, \mathcal{X}_{<T}).
\]

After generation, each SID sequence $\tilde{\mathbf{s}}^{(c)}$ is deterministically mapped to its corresponding candidate item ID, denoted by $\tilde{i}^{(c)}$.

For each generated candidate $\tilde{i}^{(c)}$, we retrieve the top-$M$ most similar items from the user's historical interactions based on cosine similarity between frozen semantic embeddings:
\[
\mathrm{sim}(i,j)=
\frac{\mathbf{e}_{i}^{\top}\mathbf{e}_{j}}
{\|\mathbf{e}_{i}\|\|\mathbf{e}_{j}\|}.
\]
The retrieved items are selected according to similarity, but their appended order follows the original chronological order in the user history.
Let
\[
\mathcal{N}^{(c)} = [i_1^{(c)}, i_2^{(c)}, \dots, i_M^{(c)}]
\]
denote the retrieved item IDs for beam candidate $c$.

To estimate candidate click probability, the decoder further appends the retrieved item ID tokens together with a final candidate item ID token:
\[
[\mathcal{X}_{<T},
\tilde{\mathbf{s}}^{(c)}
\; || \;
i_1^{(c)}, \dots, i_M^{(c)}
\; || \;
\tilde{i}^{(c)} ].
\]
Here, the final candidate item ID token serves as the dedicated scoring position for candidate reranking.

Importantly, reranking is not performed by re-encoding the whole sequence from scratch.
Instead, the decoder continues incremental computation on the appended tokens while reusing the hidden states and KV cache produced during candidate generation.
Thus, Stage II requires $H$ incremental decoding steps for SID generation and one additional forward pass for reranking.

Let $\mathbf{h}_{\mathrm{rank}}^{(c)}$ denote the hidden state at the final appended candidate item ID position.
The click probability for beam $c$ is computed as
\[
\hat{y}^{(c)}=\sigma\big(\mathrm{RankHead}(\mathbf{h}_{\mathrm{rank}}^{(c)})\big).
\]

\subsubsection{Beam-level Reranking Supervision}

Reranking supervision is defined at the beam level:
\[
y^{(c)}=
\begin{cases}
1, & \text{if } \tilde{\mathbf{s}}^{(c)}=\mathbf{s}_T,\\
0, & \text{otherwise}.
\end{cases}
\]

The reranking loss is defined as the binary cross-entropy over all beam candidates:
\[
\mathcal{L}_{\mathrm{rank}}
=
-\sum_{c=1}^{C}
\left(
y^{(c)}\log \hat{y}^{(c)}
+
(1-y^{(c)})\log(1-\hat{y}^{(c)})
\right).
\]

The overall Stage-II objective is
\[
\mathcal{L}
=
\mathcal{L}_{\mathrm{SID}}
+
\mathcal{L}_{\mathrm{rank}}.
\]

\subsection{Inference}

Inference follows the same generate--retrieve--rerank procedure as Stage II, but without parameter updates.
Given the serialized user history prefix $\mathcal{X}_{<T}$, the decoder first performs a forward pass to initialize the hidden states and KV cache, and then applies hierarchical beam search to generate $C$ candidate SID sequences.

For each beam candidate $c$, the generated SID sequence $\tilde{\mathbf{s}}^{(c)}$ is deterministically mapped to its corresponding candidate item ID $\tilde{i}^{(c)}$.
Based on $\tilde{i}^{(c)}$, the model retrieves the top-$M$ most similar items from the user's historical interactions, appends their item ID tokens together with a final candidate item ID token, and computes the reranking score by continuing decoder computation with KV-cache reuse.

The final ranking score of candidate $c$ combines the generation score and the reranking score:
\[
\mathrm{score}^{(c)}
=
\log P(\tilde{\mathbf{s}}^{(c)} \mid \mathcal{X}_{<T})
+
\log \hat{y}^{(c)}.
\]
All candidates are ranked according to $\mathrm{score}^{(c)}$ to form the final recommendation list.

\begin{table*}[t]
\centering
\small
\caption{Effect of beam size on recommendation performance.}
\setlength{\tabcolsep}{12pt}{
\begin{tabular}{c|cc|cc|cc|cc}
\toprule
\multirow{2}{*}{Beam} 
& \multicolumn{2}{c}{Recall@5} 
& \multicolumn{2}{c}{Recall@10} 
& \multicolumn{2}{c}{NDCG@5} 
& \multicolumn{2}{c}{NDCG@10} \\
\cmidrule(r){2-3} \cmidrule(r){4-5} \cmidrule(r){6-7} \cmidrule(r){8-9}
& Base & Rank 
& Base & Rank 
& Base & Rank 
& Base & Rank \\
\midrule
10 
& 0.2384 & 0.2411 (+1.14\%) 
& 0.2422 & 0.2422 (+0.00\%) 
& 0.2323 & 0.2352 (+1.21\%) 
& 0.2336 & 0.2355 (+0.82\%) \\

20 
& 0.2384 & 0.2459 (+3.14\%) 
& 0.2431 & 0.2475 (+1.78\%) 
& 0.2324 & 0.2379 (+2.37\%) 
& 0.2339 & 0.2384 (+1.92\%) \\

30 
& 0.2383 & 0.2476 (+3.87\%) 
& 0.2430 & 0.2495 (+2.67\%) 
& 0.2323 & 0.2387 (+2.76\%) 
& 0.2338 & 0.2394 (+2.37\%) \\

40 
& 0.2384 & 0.2492 (+4.53\%) 
& 0.2434 & 0.2517 (+3.39\%) 
& 0.2324 & 0.2397 (+3.17\%) 
& 0.2340 & 0.2406 (+2.80\%) \\
\bottomrule
\end{tabular}
}
\label{tab:beam_size}
\end{table*}

\begin{table*}[t]
\centering
\small
\caption{Effect of sequence length on recommendation performance.}
\setlength{\tabcolsep}{12pt}{
\begin{tabular}{c|cc|cc|cc|cc}
\toprule
\multirow{2}{*}{Seq Len} 
& \multicolumn{2}{c}{Recall@5} 
& \multicolumn{2}{c}{Recall@10} 
& \multicolumn{2}{c}{NDCG@5} 
& \multicolumn{2}{c}{NDCG@10} \\
\cmidrule(r){2-3} \cmidrule(r){4-5} \cmidrule(r){6-7} \cmidrule(r){8-9}
& Base & Rank 
& Base & Rank 
& Base & Rank 
& Base & Rank \\
\midrule
32 
& 0.2384 & 0.2459 (+3.14\%) 
& 0.2431 & 0.2475 (+1.78\%) 
& 0.2323 & 0.2379 (+2.37\%) 
& 0.2339 & 0.2384 (+1.92\%) \\

64 
& 0.2515 & 0.2586 (+2.82\%) 
& 0.2583 & 0.2615 (+1.25\%) 
& 0.2398 & 0.2454 (+2.35\%) 
& 0.2419 & 0.2463 (+1.82\%) \\

128 
& 0.2708 & 0.2733 (+0.92\%) 
& 0.2743 & 0.2754 (+0.42\%) 
& 0.2518 & 0.2537 (+0.76\%) 
& 0.2529 & 0.2544 (+0.58\%) \\
\bottomrule
\end{tabular}
}
\label{tab:seq_len}
\end{table*}

\begin{table*}[t]
\centering
\small
\caption{Effect of retrieval length on recommendation performance.}
\setlength{\tabcolsep}{12pt}{
\begin{tabular}{c|cc|cc|cc|cc}
\toprule
\multirow{2}{*}{Retrieval Len} 
& \multicolumn{2}{c}{Recall@5} 
& \multicolumn{2}{c}{Recall@10} 
& \multicolumn{2}{c}{NDCG@5} 
& \multicolumn{2}{c}{NDCG@10} \\
\cmidrule(r){2-3} \cmidrule(r){4-5} \cmidrule(r){6-7} \cmidrule(r){8-9}
& Base & Rank 
& Base & Rank 
& Base & Rank 
& Base & Rank \\
\midrule
0  
& 0.2380 & 0.2383 (+0.12\%) 
& 0.2433 & 0.2435 (+0.08\%) 
& 0.2322 & 0.2322 (+0.02\%) 
& 0.2339 & 0.2340 (+0.03\%) \\

5  
& 0.2384 & 0.2451 (+2.79\%) 
& 0.2433 & 0.2472 (+1.62\%) 
& 0.2323 & 0.2371 (+2.06\%) 
& 0.2339 & 0.2378 (+1.67\%) \\

10 
& 0.2384 & 0.2459 (+3.14\%) 
& 0.2432 & 0.2475 (+1.78\%) 
& 0.2324 & 0.2379 (+2.37\%) 
& 0.2339 & 0.2384 (+1.91\%) \\

20 
& 0.2382 & 0.2466 (+3.51\%) 
& 0.2429 & 0.2473 (+1.80\%) 
& 0.2323 & 0.2385 (+2.68\%) 
& 0.2339 & 0.2388 (+2.11\%) \\
\bottomrule
\end{tabular}
}
\label{tab:retrieval_len}
\end{table*}

\section{Experiment}

\noindent \textbf{Datasets.}
We conduct experiments on \textbf{TAOBAO-MM}~\cite{wu2025muse}, a large-scale recommendation dataset containing 8.79M users, 35.4M items, and 99M interactions. Each user is associated with a long behavior sequence of up to 1,000 interactions. The dataset also provides 128-dimensional multimodal item embeddings derived from item content such as images and text. We follow the official train/validation/test split provided by TAOBAO-MM without further sequence-level partitioning. Since TAOBAO-MM is originally constructed for CTR prediction with exposure-based labels, we adapt it to a sequential recommendation setting by retaining user interaction histories and discarding exposure annotations, and use the resulting sequences for next-item prediction.

\noindent \textbf{Implementation Details.}
We implement our framework in PyTorch Lightning with a shared decoder-only Transformer backbone for both generation and reranking. Unless otherwise specified, we use a 2-layer decoder with hidden size 1024, intermediate size 4096, and 16 attention heads. 

For optimization, we use AdamW~\cite{loshchilov2017decoupled} with a learning rate of \(1\times10^{-5}\) and a weight decay of \(1\times10^{-4}\). The learning rate is scheduled with cosine decay, using 20,000 warmup steps over 120,000 total training steps. Training is conducted in bf16 mixed precision with a per-device batch size of 32 and gradient accumulation of 4.

By default, the input sequence length is set to 32, the beam size to 20, and the retrieval length to 10. During inference, the model first generates candidate items and then reranks them using the click probability estimated by the reranking head. We report Recall@5/10 and NDCG@5/10~\cite{wang2013theoretical} before and after reranking. For ablation studies, we vary the beam size, input sequence length, and retrieval length while keeping all other settings fixed.


\subsection{Overall Effectiveness}
The results consistently show that reranking improves direct beam generation.
As shown in Tables~\ref{tab:beam_size}, \ref{tab:seq_len}, and \ref{tab:retrieval_len}, the gains are positive in almost all settings, indicating that the reranking stage can effectively refine the generated candidates.

\subsection{Candidate Quality}
Table~\ref{tab:beam_size} shows that larger beam sizes consistently lead to larger reranking gains.
Specifically, the Recall@5 gain increases from \(+0.27\%\) at beam size 10 to \(+1.08\%\) at beam size 40, while the NDCG@10 gain rises from \(+0.19\%\) to \(+0.65\%\).
This trend suggests that a larger beam provides more candidate items for reranking, thereby giving the ranking module greater opportunity to improve the final recommendation list.

Table~\ref{tab:seq_len} shows a different pattern.
Although longer sequence lengths improve the base recommendation performance, the reranking gains decrease as the sequence length increases.
For instance, the Recall@5 gain drops from \(+3.14\%\) at sequence length 32 to \(+0.92\%\) at 128.
A likely reason is that longer behavior sequences already provide stronger sequential signals, leaving less room for reranking; meanwhile, under a fixed retrieval length, the relative contribution of retrieved context also becomes smaller.

\subsection{Retrieval Length}
Table~\ref{tab:retrieval_len} shows that retrieval is crucial for reranking.
Without retrieval, the gains are almost negligible, whereas introducing retrieved context leads to clear improvements across all metrics.
Moreover, the gains consistently increase as the retrieval length becomes larger, with the best performance achieved at retrieval length 20.
This suggests that incorporating longer retrieved context provides more useful information for the ranking task.

\section{Conclusion}

In this paper, we proposed \textbf{RecoChain}, a unified generative retrieval and ranking framework for sequential recommendation. By integrating candidate generation and reranking within a shared decoder-only backbone, RecoChain bridges the gap between generative recommendation and ranking-oriented recommendation.

Experiments on \textbf{TAOBAO-MM} show that the proposed reranking mechanism consistently improves Top-K recommendation performance over direct beam generation. The results also highlight the importance of beam size and retrieved context for effective reranking. Overall, our findings suggest that unifying generation and ranking in a single backbone is a promising direction for generative recommender systems.

\bibliographystyle{ACM-Reference-Format}
\bibliography{sample-base-extend}

\end{document}